\documentclass[epj,final]{svjour}
\usepackage{graphicx,amstext}
\renewcommand{\epsilon}{\varepsilon}
\renewcommand{\phi}{\varphi}
\newcommand{\vs}{{\it vs.\ }}
\newcommand{\al}{{\it et al.\ }}
\newcommand{\ie}{{\it i.e.\ }}
\newcommand{\eg}{{\it e.g.\ }}
\newcommand{\pff}{(TMTSF)$_2$PF$_6$}
\newcommand{\Aff}{(TMTSF)$_2$AsF$_6$}

\begin{document} 
\sloppy 
\title{Coexistence of 
Superconductivity  and Spin Density  Wave orderings in the 
organic superconductor (TMTSF)$_2$PF$_6$} 
\titlerunning{SC and SDW coexistence in (TMTSF)$_2$PF$_6$} 
\author{T.~Vuleti\'{c}\inst{1,2}\and 
P.~Auban-Senzier\inst{1} \and C.~Pasquier\inst{1}\and 
S.~Tomi\'{c}\inst{2} \and D.~J\'{e}rome\inst{1} \and 
M.~H\'eritier\inst{1} \and K.~Bechgaard\inst{3} 
\mail{tvuletic@ifs.hr}} 
%
\institute{Laboratoire de Physique des Solides, 
Universit\'{e}Paris-Sud, F-91405 Orsay, France \and 
Institut za fiziku, p.p. 304, HR-10001 Zagreb, Croatia 
\and Dept.~of Condensed Matter Physics and Chemistry, 
Ris\o~National Laboratory, DK-4000 Roskilde, Denmark} 
\date{Received: date / Revised version: date} 
\abstract{ 
The  phase diagram of the organic superconductor \pff~has 
been revisited  using transport measurements with an 
improved control of the applied pressure. We have found a 
0.8 kbar wide pressure domain below the critical point 
(9.43~kbar, 1.2 K) for the stabilisation of the 
superconducting ground state featuring a  coexistence regime 
between spin density wave (SDW) and superconductivity (SC). 
The inhomogeneous character of the said pressure domain is 
supported by the analysis of the resistivity between 
$T_{\text {SDW}}$ and $T_{\text {SC}}$ and  the 
superconducting critical current. The onset temperature 
$T_{\text {SC}}$ is practically constant ($1.20 \pm 0.01$ K) 
in this region where only the SC/SDW domain proportion below 
$T_{\text {SC}}$ is increasing under pressure. An 
homogeneous superconducting state is recovered  above the 
critical pressure with $T_{\text {SC}}$ falling at 
increasing pressure. We propose a model comparing the free 
energy of a phase exhibiting a segregation between SDW and 
SC domains and the free energy of homogeneous phases which 
explains fairly well our experimental findings. 
\PACS{ 
{71.10.Ay} {Fermi-liquid theory and other phenomenological models} 
{74.25.Dw} {Superconductivity phase diagrams} 
{74.70.Kn} {Organic superconductors} 
{75.30.Fv} {Spin-density waves} 
}
}

\maketitle

\section{Introduction} 
The question of competition between a 
magnetic state and a superconducting (or even only metallic) 
state is one of the key actual features in the domain of 
strongly correlated electrons. This point is extensively 
studied in lamellar superconductors such as the high-T$_c$ 
copper oxide superconductors \cite{orenstein00} and heavy fermion 
compounds \cite{mathur98}. Such a competition is also common in 
organic materials and an 
antiferromagnetism(AF)-superconductivity(SC) coexistence 
region has been observed recently in the 
$\kappa$-(BEDT-TTF)$_{2}$X compounds under well controlled 
pressure \cite{lefebvre00,Ito96}. An even better  characterised systems
featuring the competition of AF(or SDW) and metal or superconductivity is
formed by the quasi-one dimensional organic superconductors of the
(TM)$_{2}$X family, \ie the Bechgaard-Fabre salts. Even after 25 years of
investigation this competition and the exact transition from magnetism to
superconductivity remains unclear
\cite{jerome91,ishiguro98,bourbonnais99,bourbonnais99a}.  In particular, the
prototype and most studied compound, \pff, has an ideal position in the
pressure-temperature phase diagram and an improved experimental setup allows
now for investigation of this transition.

In the framework  of the Fermi liquid picture, which is 
expected to be valid in this low temperature part of the 
phase diagram, it is well known that the formation of a SDW 
phase in the Bechgaard salts is due to a large Kohn anomaly, 
associated to an almost perfect nesting of the metallic 
Fermi surface.\ When a pressure is applied, the nesting 
properties are spoiled, which causes a decrease of the 
ordering temperature $T_{\text {SDW}}$, up to a critical 
pressure $p_{c},$ at which $T_{\text {SDW}}$ vanishes as 
observed in the title compound where $T_{\text SDW}=12$~K  
at ambient pressure and becomes a superconductor under high 
pressures with a critical temperature in the Kelvin range 
\cite{jerome80}. However, it has been noticed in the first 
experimental reproduction of organic superconductivity under 
pressure \cite{greene80} that both SC and a metal-insulator 
transition appear to coexist in a narrow pressure domain 
upon cooling. Such investigation  of the phase diagram of 
the parent compound (TMTSF)$_2$AsF$_6$ by Brusetti \al 
\cite{brus82}  has led to similar observations. Then we may 
add that such competition is also observed in 
(TMTSF)$_2$ClO$_4$ \cite{tomic82} where anion disorder is 
now the driving parameter. Using EPR at low field with 
helium gas pressure techniques in \pff~under pressure, 
Azevedo \al \cite{azevedo84} have reported the existence of a transition 
between a SDW phase and a superconducting one on cooling in 
a narrow pressure domain (5.5-5.7 kbar). However, they 
claimed to be able to rule out the possibility that some 
parts of the sample are in the SDW state and other in the 
superconducting state.

On a theoretical point of view, the merging of SDW into the 
SC state around a critical pressure has been extensively 
studied. The precursor work of Schulz \al \cite{schulz81} 
studying the border between SDW and SC states in \pff~has 
suggested two possible scenarios, namely, the existence of a 
quantum critical point between SDW and SC or a first order 
transition line between the insulating and superconducting 
states.  On the other hand, it is well known that, at a 
pressure lower than the critical pressure, but close to it, 
a ``semi-metallic'' SDW phase is formed, with small pockets 
of unpaired charge carriers, because the SDW gap is not 
opened on the whole Fermi surface\cite{Yamaji0}. In some 
sense, there is a coexistence of metallic and magnetic 
phases coming from different parts of the reciprocal space.
The possibility of a microscopic coexistence of 
superconducting and SDW phases, at lower temperature, has 
been studied in details\cite{Yamaji2}, \cite{Hasegawa}, \cite{Yamaji3}. The conclusion was negative, 
because the density of states of the unpaired carrier 
pockets, left by the SDW ordering, is strongly reduced 
compared to that of the metallic phase.\ Such a reduction 
drastically decreases the effective superconducting coupling 
$gN\left( 0\right)$, leading to an exponentially small critical temperature.For the sake of completeness
we can recall that Machida  has proposed a model of microscopically competing
orders coming from the different parts of the Fermi surface 
leading in turn to a coexistence of both orders at low
temperature \cite{Machida81}.

In this article, using an optimised pressure control, we 
reinvestigate the critical region of the phase diagram of 
\pff, which features the phase boundary between the spin 
density wave and the superconducting state. Studying the 
temperature dependence of the resistance and the 
superconducting critical current, we can obtain some 
evidence for the existence of an inhomogeneous phase 
forming in the vicinity of the border. We also propose a 
simple model which demonstrates that a phase segregation 
SDW/SC, with formation of macroscopic domains, is always 
favourable compared to the homogeneous phases.

\section{Experimental} 
We worked with a nominally pure \pff~single crystal 
originating from the same batch of high quality crystals 
used in a previous study \cite{vuletic01}. The crystal had 
standard needle shape and dimensions $3\times 0.2\times  0.1\; mm^{3}$.The four annular contact
geometry was used: gold was evaporated on the sample and the leads were attached with silver
paste. The contact resistances were just 2-3 Ohms. The high quality of the crystal was
confirmed  by the resistivity ratio ($\rho_{300K}/\rho _{T_{\text  SDW}}$) of the order of
1000. All electrical measurements 
were performed along the needle \textbf{a}-axis. Linear 
resistance measurements were performed using  a standard AC 
low frequency technique. High critical currents measurements 
were performed using a DC pulsed technique described 
elsewhere \cite{vuletic01} with 10~$\mu$s short pulses and 
amplitudes up to 100~mA. 
The pulse repetition period was 40 ms, \textit{i.e.} 4000 times longer than the pulse. This is
enough to avoid Joule heating of the sample even at the highest amplitudes of 100 mA.
Non-heating was also checked by the shape of the pulse displayed on the osciloscope. If heating
effects were present at the highest currents (0.3, 1 mA)  then the SDW
resistance just above $T_{sc}$ should get lower at increasing current. Of course, as one
can find in the Fig.5, there is no difference in resistances measured in the SDW phase using
currents from 0.001 to 1 mA.

The pressure cell was then plugged 
on a Helium3 cryostat capable of reaching 0.35~K.

The pressure was applied in a regular beryllium-copper cell, 
with silicon oil inside a Teflon cup as the pressure 
transmitting medium. This liquid does not freeze abruptly 
but solidifies continuously thus reducing mechanical 
stresses and pressure shifts which are common effects at the 
freezing points of other liquids. This allowed numerous 
thermal cyclings of the same sample without noticeable 
cracks.

However, the drawback of such a technique is the requirement 
to change pressure only when the cell is warmed up to room 
temperature. For an accurate pressure determination, we used 
an InSb pressure gauge\cite{koncz78}, located inside the 
cell close to the sample. The informations from the pressure 
gauge can be also cross-checked by the known phase diagram 
of \pff~\cite{cha12}. The InSb gauge was calibrated at 
ambient temperature against a manganine gauge establishing a 
linear pressure dependence of the resistance at a rate of 
2.5\%/kbar in the 6--12 kbar range.

The measurements were conducted in 19 consecutive runs, as 
follows (see Table~\ref{tablica1}). For all runs, after 
application of the pressure at room temperature, the cell 
was then immediately cooled down to 0.35 K. The electrical 
measurements done, the sample was warmed back to room 
temperature  and a subsequent increase or decrease of 
pressure was immediately applied before cooling again. The 
temperature sweep rates in cooling and warming did not 
exceed $\pm$60 K/h. The change  of pressure was checked  by 
the resistance of the InSb gauge: for instance, a 0.8\%  
increase  in resistance determined a 300 bar increment in 
pressure (the simultaneous decrease of the resistance at 
room temperature was also used as a secondary pressure 
gauge). Additionally, the pressure coefficient of the InSb 
resistance measured at ambient temperature was equal to that 
measured at 6 K. Thus, we confirmed that all pressure steps 
we made were measured by the InSb gauge with an accuracy of 
$\pm$30 bar.

Starting in run \#1, at a pressure $p_{1}=6.8$~kbar, we were 
able to sweep a 4 kbar wide region with increments ranging 
from 300 to 100 bar. From run \#1 to run 
\#16, pressure was always increased in this manner except a 
1400 bar increment for the last run. Then from run \#16 to 
run \#17, we made a large pressure drop of 2350$\pm$50 bar. 
This value was calculated from the InSb resistance and also 
later confirmed by the phase diagram of \pff. That is, the R 
\vs T curve for this run \#17 was almost identical to the 
curve of the run \#10 (both SDW and SC critical temperatures 
were found identical within experimental errors). Small 
pressure increments for run \#18 and \#19 were again 
performed in the usual manner. This procedure allowed us to 
investigate the pressure domain 8.65--9.3 kbar with a 
control of the pressure which could not be achieved in the 
early studies.

\begin{table}
\begin{tabular}{|r|l|l|l|l|l|l|l|}
\hline
RUN&pressure&$T_{\text {SDW}}$&$T_{\text {SC}}$&$\Delta$&$R_{\infty}$&$c$\\
&(kbar)&(K)&(K)&(K)&(m$\Omega$)&vol\%\\
\hline
\hline 1&6.8&7.9&&9.38&12.1&0.03\\
\hline 2&7.1&7.1&&&&\\  
\hline 3&7.3&6.6&&&&\\
\hline 4&7.5&6.2&&&&\\  
\hline 5&7.7&5.9&&&&\\  
\hline 6&7.85&5.6&&&&\\
\hline 7&8.15&5.3&&	6.35&12.0&0.09\\ 
\hline 8&8.35&4.7&&&&\\
\hline 9&8.45&4.4&&5.74&11.30&0.2\\ 
\hline 10&8.65&3.8&1.18&5.56&10.60&0.3\\
\hline 17&8.65&3.8&1.18&&&\\
\hline 11&8.75&3.5&1.19&5.58&8.772&0.36\\
\hline 18&8.9&2.85&1.21&6.90&2.473&1.25\\
\hline 12&8.95&2.85&1.21&6.70&2.146&1.4\\
\hline 13&9.1&2.45&1.21&6.90&0.320&5\\
\hline 19&9.2&1.8&1.21&6.30&0.0452&17\\
\hline 14&9.3&1.4&1.21&6.80&0.0022&89\\
\hline 15&9.6&&1.195&&&\\
\hline 16&11&&1.106&&&100\\
\hline
\end{tabular}

\caption{The complete range of pressures applied on the 
single crystal of \pff~studied in this work. Number of 
the run gives the chronological order of the experiment. 
Observed spin-density wave, $T_{\text {SDW}}$ and 
superconducting, $T_{\text {SC}}$, transition temperatures 
are given. The activation energy, $\Delta$, asymptotic 
resistances, $R_{\infty}$ and the volume proportion 
$c$~(vol\%) of the metallic phase when the sample is in the 
coexistence regime are obtained from the fits of $R$ \vs~$T$ 
data to the recalculated Arrhenius law, Eq.~\ref{rhoT}.} 
\label{tablica1} 
\end{table}

\section{Experimental observations} 

\subsection{The metallic state : $T>T_{\text {SDW}}$, $T>T_{\text {SC}}$} 
\begin{figure}
\centering\resizebox{0.44\textwidth}{!}{\includegraphics*{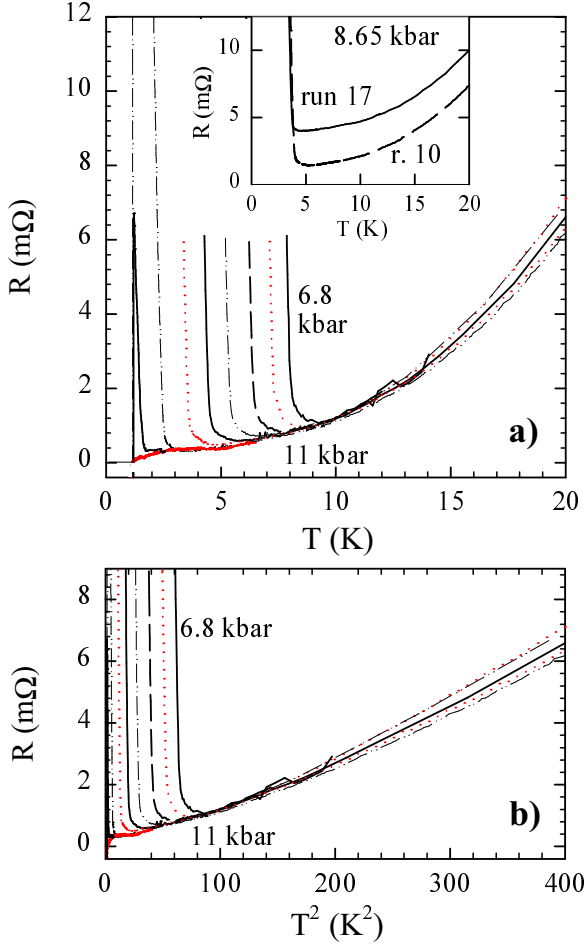}}
\caption{ (a) Resistance \vs temperature curves for the usual set
of pressures (see legend in Fig.~\ref{Plot2}), the sharpness of the transitions may be fully
appreciated. The behaviour above M - SDW transition is
characterised by the unexpected positive pressure
coefficient. This is due to the residual resistance, which
increases for consecutive pressure runs.  In the inset: A
detail of the minimum in the $R$~\vs~$T$ curves. For runs 10
and 17, both measured at 8.65 kbar, resistance shift
amounted to 2.5 m$\Omega$.
 (b) Resistance \vs temperature
squared curves for the given list of pressures. Low
temperature behaviour approaches the $T^2$ law of 3D metals.}
\label{Plot1}
\end{figure}

In Fig.~\ref{Plot1}a, we show the 
resistance \vs temperature below 20 K for a set of 
characteristic pressures from 6.8 to 9.2 kbar together with 
the data at 11 kbar where a direct metal-SC transition is 
observed. The temperature dependence of the resistance in 
the metallic state (we concentrate on temperatures below 20 
K) is quadratic, as expected when the electron scattering is 
dominated by electron-electron interactions\cite{gorkov96}.

We have noticed a shift of the $R$ \vs~$T$ curves by a 
temperature independent resistance value after each pressure 
run, without any change of the actual temperature 
dependence. The resistance curves are usually shifted 
0.1--0.2 m$\Omega$ upwards after each run (see caption of 
Fig~\ref{Plot1}a). This effect is most clearly seen in the 
metallic state below 20 K since the absolute values of the 
resistance (1--2 m$\Omega$) become then comparable to the 
offset. An example of this offset is shown in 
Fig~\ref{Plot1}a. The two resistance curves are for run \#10 
and \#17, \ie they were measured at equal pressures, but 
with 7 runs performed in-between. The behaviour of run\#17 
can be made equal to the run \#10 provided an offset of 2.5 
m$\Omega$ is subtracted from the resistance values of the 
earlier run \#10. We tend to relate this effect to an 
increase of the residual resistance due to the cumulative 
creation of defects after each temperature cycle.  We will 
show later that such defects are not cracks as they would 
add junctions in the sample, which will be easy to detect in 
the superconducting state. The added defects, then, could be 
of point disorder nature of unknown origin. It should be 
noted that such phenomenon was previously unreported. Still, 
it can not be excluded in all previous studies, since this 
extensive thermal cyclings were not performed before. 

However, the metallic state resistance was the only aspect 
influenced by the extensive thermal cycling of the sample. 
The values of resistance in the SDW state which are orders 
of magnitude higher were not influenced by the thermal 
cycling and we could not notice any influence on the 
determination of the transition temperatures. Since a plot 
of the resistance \vs $T^{2}$ (see Fig.~\ref{Plot1}b) 
reveals the existence of a quadratic temperature dependence 
below about 12 K with a residual resistance increasing 
slightly after each pressure run, we have decided to use the 
value of the residual resistance obtained in run \#1 as the 
reference value, assuming that the sample was the least 
damaged in the first run, as compared to all subsequent 
runs. So doing, we noticed that the resistance is weakly 
pressure dependent at 20 K and becomes insensitive to 
pressure below 10 K in the pressure domain investigated.

\begin{figure}
\centering\resizebox{0.46\textwidth}{!}{\includegraphics*{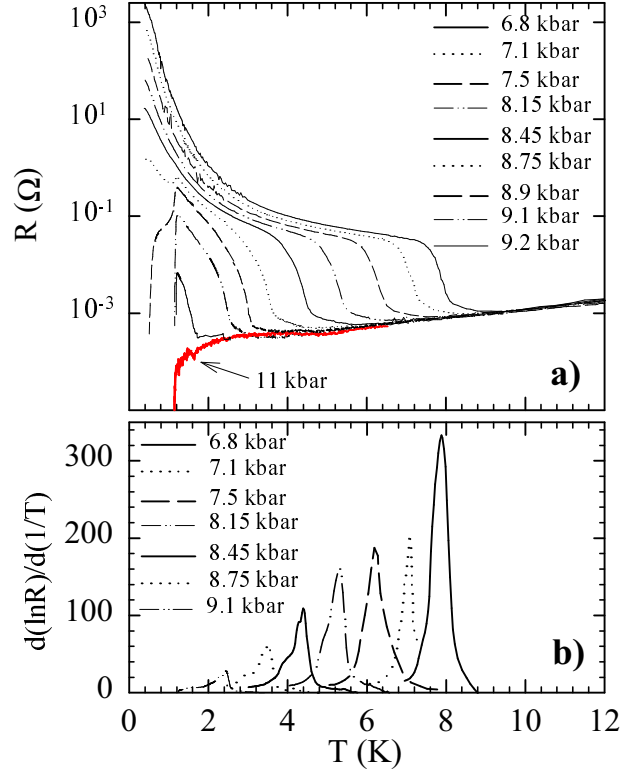}}
\caption{(a) Logarithm of resistance
\vs temperature  in cooling, at different pressures. Above
the SDW transition (step-like features) the resistance
decreases as expected for a metal.  (b) $T_{\text {SDW}}$ is
defined as the peak in the derivative of logarithm of
resistance over inverse temperature, $\partial(\ln
R)/\partial(1/T)$ \vs temperature, $T$. Peaks are shown for the pressures
denoted in the figure above.}
\label{Plot2}
\end{figure}

\subsection{The SDW regime: $p<8.6$~kbar, $T<T_{\text 
{SDW}}$} 

\begin{figure}
\centering\resizebox{0.46\textwidth}{!}{\includegraphics*{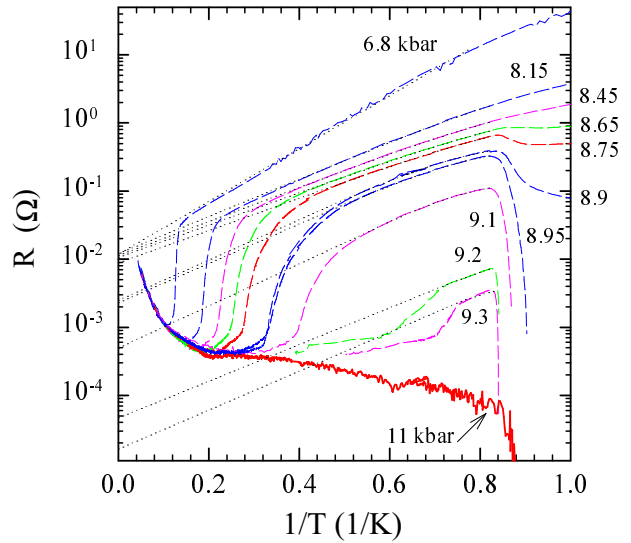}}
\caption{Logarithm of the resistance \vs inverse temperature for the given range of pressures, depicting behaviour in the SDW phase. Fits to the recalculated Arrhenius law (dotted straight lines, Eq.~\ref{rhoT}) give the same asymptotic resistances, $R_{\infty}$ for the curves inside the pure SDW region. Inside the SDW/M region asymptotic resistances decrease by several orders of magnitude.}
\label{Plot3}
\end{figure}

As the SDW regime, we denote  the low pressure, low 
temperature region  where the transition to the SDW state is 
observed as a sharp increase of the resistance 
(Fig.~\ref{Plot2}a) without any hysteretic behaviour when 
sweeping temperature up and down.  The exact point of the 
transition is defined as the temperature of the maximum of 
the logarithmic derivative of the resistance with respect to 
the inverse temperature, $\partial(\ln R)/\partial(1/T)$ 
\vs~$T$. In Fig.~\ref{Plot2}b, we show these maxima 
corresponding to the $R$~\vs~$T$ curves above. Activation 
energy values, $\Delta $, are well defined inside broad 
temperature spans, especially for the pressures in this 
regime, 6.8--8.45 kbar (Fig.~\ref{Plot3}). We note here that 
the asymptotic resistance at infinite temperature, 
$R_{\infty}$, remains constant in this pressure range as 
expected in a standard semiconducting model.

\subsection{The SC regime: $p>9.43$ kbar, $T<T_{\text {SC}}$}

This pressure domain has been briefly investigated  
at two subsequently applied pressures (9.6 and 11 kbar). A 
direct transition from a metallic to a superconducting state 
was observed. $T_{\text {SC}}$ was defined  as the onset of 
the resistance drop. These two pressures differ only in 
$T_{\text {SC}}$, which are $1.195\pm0.005$ K and 
$1.106\pm0.005$ K respectively. The transitions are very 
narrow (30-40 mK) and no hysteresis has been observed at 
$T_{\text {SC}}$. The remaining shift of less than 3 mK 
observed between cooling and warming curves can be 
attributed to the finite speed of the temperature sweep, 
$\pm1$ mK/s and the thermal inertia of the pressure cell.
The mean pressure dependence  $\partial T_{\text 
{SC}}/\partial p \approx -0.07$~K/kbar is in agreement with 
the value already reported by Schulz~\al~\cite{schulz81}.

\subsection{The inhomogeneous SDW regime: $8.6<p<9.43$~kbar}

We denote as the inhomogeneous regime phase space where both 
SDW and M (eventually SC) states clearly manifest. Taking 
advantage of the good pressure control, we managed to 
investigate eight pressure points with the same sample in 
this narrow pressure range. They were measured in two groups 
of five and of three consecutive runs (see 
Table~\ref{tablica1}). The second group was measured after a 
large pressure decrease. This decrease was precisely 
targeted to reproduce a point (8.65 kbar) in the lower end of the pressure range of interest. This provided a check for 
our capability of controlling accurately the pressure and an 
opportunity to investigate in more detail this pressure 
range in the usual manner by small pressure increments. As 
already noted, runs \#10 and \#17 gave almost identical 
$R$~\vs~$T$ curves except for the residual resistance 
offset.

\subsubsection{$8.6<p<9.43$~kbar, $T_{\text {SC}}<T<T_{\text {SDW}}$: SDW/Metal}

\begin{figure}
\centering\resizebox{0.46\textwidth}{!}{\includegraphics*{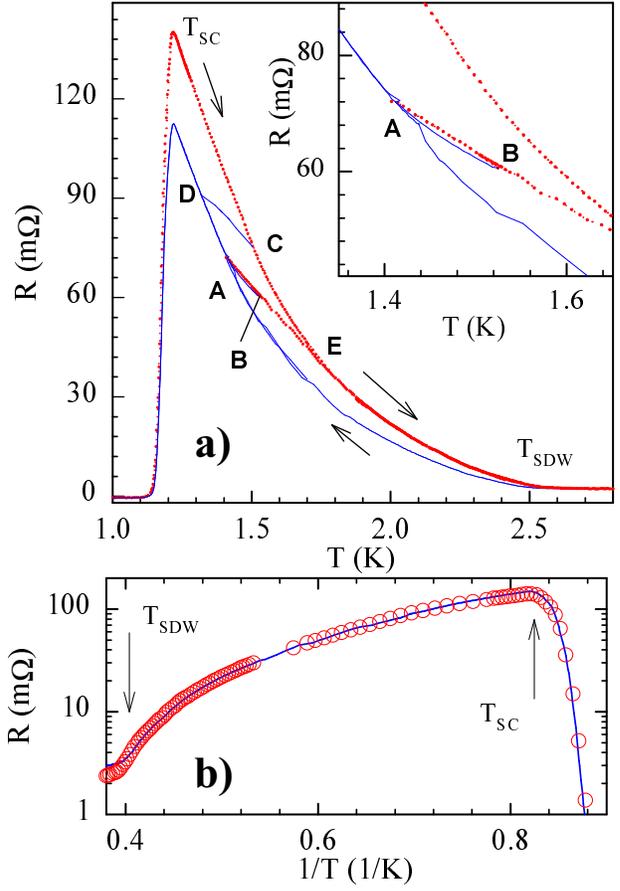}}

\caption{a)Resistance \vs temperature at 9.1 kbar, inside the coexistence range. Solid and dotted lines denote cooling and warming, respectively.
The extremal hysteretic loop is recorded when the temperature sweep starts from above $T_{\text {SDW}}$, then cooling continues through points A, D, through $T_{\text {SC}}$, and into superconducting state. Below at least 1 K the sweep may be reversed and warming proceeds through $T_{\text {SC}}$, C, E and back above  $T_{\text {SDW}}$ into the SDW phase. Different hysteretic loops appear when temperature sweep is reversed between
$T_{\text {SC}}$ and $T_{\text {SDW}}$.
Loop 1: cooled from above $T_{\text {SDW}}$ to {\bf A}, reversed, warmed to {\bf E} and further to above $T_{\text {SDW}}$.
Loop 2: warmed from below $T_{\text {SC}}$ to {\bf C}, reversed, cooled to {\bf D} and further to below $T_{\text {SC}}$.
Loop 3: cooled from above $T_{\text {SDW}}$ to {\bf A}, reversed, warmed to {\bf B}, reversed, warmed to {\bf A} and further to below $T_{\text {SC}}$.
Inset: The position of point {\bf B} is shown.
b) Corresponding $\log R$ \vs $1/T$ plot of normalised extremal
 curves in cooling (solid line) and in warming (open
points).}
\label{Plot4}
\end{figure}

As shown in Fig.~\ref{Plot4}, a strong hysteretic behaviour
is observed in this pressure and temperature range between
cooling and warming resistance curves leading to the
suggestion of an inhomogeneous electronic structure.
One possibility is the observation of the famous SDW2 state
awaited just below the critical pressure (see the discussion
below), but hysteretic behaviour \textit{vide-infra} is not
reasonably compatible with this image. The second
possibility is a phase segregation with the existence of
metallic domains in a SDW background whose characteristics
(size and relative disposition) is strongly temperature
dependent.

The  extremal hysteretic resistance loop is recorded when 
the temperature sweep starts above $T_{\text {SDW}}$, 
reverses below $T_{\text {SC}}$ and ends above $T_{\text 
{SDW}}$ again. Different hysteretic loops appear when 
temperature sweep is reversed between $T_{\text SC}$ and 
$T_{\text SDW}$ (see Fig.\ref{Plot4}a for a representative 
situation at 9.1 kbar). $T_{\text SC}$ values determined 
from either cooling or warming curves are equal (Fig.~\ref{Plot4}b), within a 
few mK, as for the direct metal-SC transitions described above. 
It is quite interesting to note that a highly similar hysteretic behaviour 
of the thermopower was observed in the CDW state of (NbSe$_4$)$_{10}$I$_3$~\cite{Smont92}.

\subsubsection{$8.6<p<9.43$~kbar, $T<T_{\text {SC}}$: SDW/SC}

In this pressure range, while SDW transitions are still well 
defined, $R$~\vs~$T$ curves are characterised by a sharp 
resistance drop at $T=1.2\pm0.01$ K. We consider this 
feature to be the manifestation of the condensation of the 
free-electron domains into superconducting domains. The SC 
domains either could percolate and form large domains for 
higher pressures, either, they could link, thanks to 
sufficiently narrow weak-links allowing for Josephson effect 
between them at lower pressures. Both mechanisms lead to the 
zero resistance state.

\begin{figure} 
\centering\resizebox{0.46\textwidth}{!}{\includegraphics*{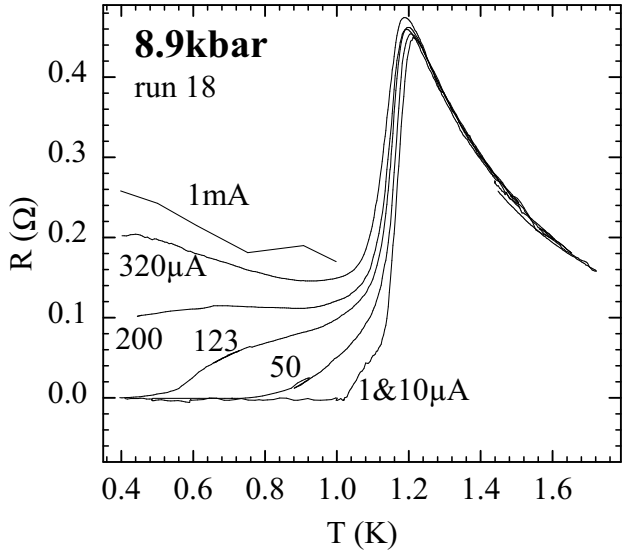}} 
\caption{Resistance \vs current at 8.9 kbar. 
The zero resistance is reached for currents as high as 123 
$\mu$A and it was not possible to completely suppress the 
resistance drop  even with a 1 mA measurement current.} 
\label{Plot5} 
\end{figure}

Initially, at $p=8.65$ kbar, the resistance drops to about 
30\%  of the resistance which would be observed if an 
Arrhenius behaviour of the SDW state was extended below 1.2 
K. This drop is only observed if small (1 to 10 $\mu$A) 
measuring currents are used. For higher currents (100 
$\mu$A), the resistance drop disappears and the usual 
Arrhenius behaviour is recovered. Then at 8.9 kbar, for the 
lowest currents, the resistance drops in fact to zero (below 
our measurable limit of 0.001 m$\Omega$ ). At 8.9 kbar the 
resistance drop is suppressed concomitantly with the 
increase of  current as shown in Fig.~\ref{Plot5}. Still, 
for an current of 1 mA, the resistance drop is far from 
being completely suppressed (it was not possible to use 
higher current in order to avoid Joule heating of the 
sample). Generally, detection of zero resistance at the 
lowest pressures inside the SDW/SC region requires the 
lowest measuring current and/or the lowest possible 
temperatures. 

\begin{figure}
\centering\resizebox{0.46\textwidth}{!}{\includegraphics*{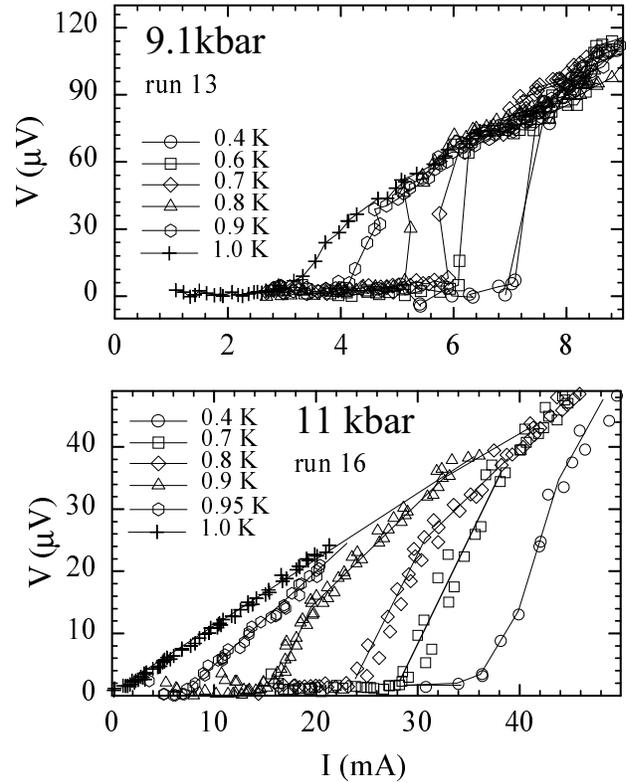}}
\caption{Voltage--current, $V$-$I$ characteristics of the sample at various
pressures. (a) At 9.1 kbar, at lowest temperature 0.4 K, the critical current
($I_c=7$~mA) is order(s) of magnitude higher than at  only 0.2 kbar
lower pressure. (b) At 11 kbar the SC state
is considered to be homogeneous and $I_c$ reaches the maximum value of 40mA.}
\label{Plot6}
\end{figure}

As pressure is further increased to 9.1--9.3 
kbar, a sharp drop of the resistance to zero within 30--40 
mK below the onset temperature is obtained regardless the 
weak amplitude of the current. In this high pressure 
regime, we used the pulse technique to determine the 
critical current, $I_{c}$ without heating effects: at 9.1 
kbar, $I_{c}=7$ mA, but at 9.3 kbar, $I_{c}$ raised up to 
30--40 mA as shown in Fig.~\ref{Plot6}. At 11 kbar, the 
critical current is of the same order of magnitude.

\section{Discussion}
\subsection{Detailed $p,T$ phase diagram of \pff}

An accurate determination of both transition 
temperature and pressure enables us to provide a precise 
$T_{\text {SDW}}$ \vs pressure line, up to the critical 
point ($p_c$=9.43 kbar, 1.2 K) where the suppression of the 
SDW instability occurs and the SC phase is fully 
established. That is, for all runs below $p_c$, we have 
observed well defined and narrow metal-SDW transitions. For 
pressures close to the critical point, the transitions are 
becoming broader as reported in references  \cite{brus82} 
and \cite{biskup95}. However, thanks to the good quality of 
our sample, this effect did not prevent the determination of 
the transition temperatures even at pressures close to the 
critical point. This is a salient result of this work 
because Brusetti~\al~\cite{brus82} and 
Bi\v{s}kup~\al~\cite{biskup95} claimed that the SDW 
transition broadens with pressure approaching the critical 
pressure. Consequently, they were unable to perform an 
accurate study of the $T_{\text {SDW}}$ \vs pressure phase 
boundary at pressures close to $p_c$.

\begin{figure}
\centering\resizebox{0.46\textwidth}{!}{\includegraphics*{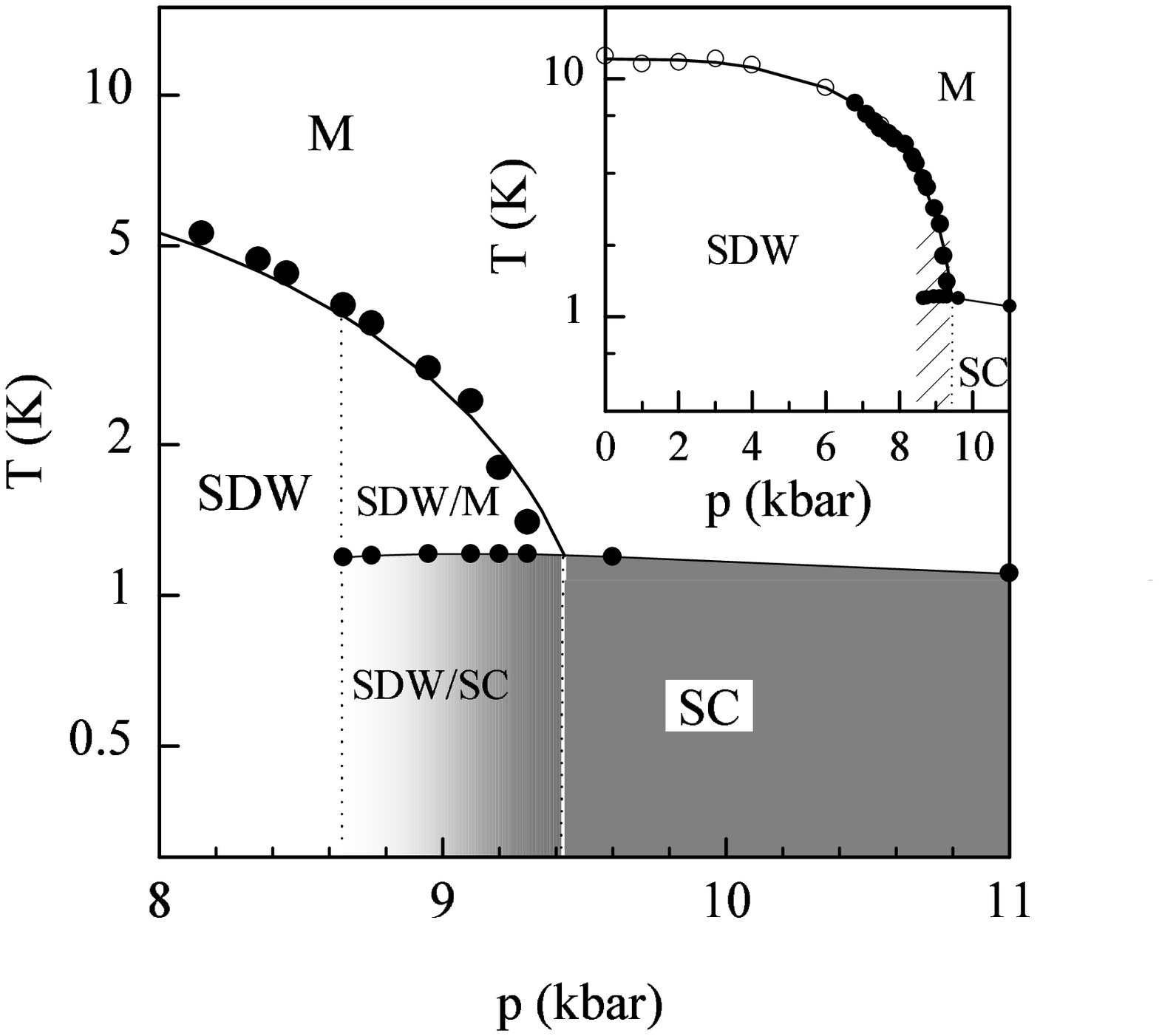}}
\caption{High pressure--low temperature phase diagram of \pff~material. SDW/M denotes the region where  metallic, M, and SDW phases coexist inhomogeneously, below $T_{\text {SDW}}$ line (large full points). Below $T_{\text SC}=1.20 \pm 0.01$ K line (small full points), this coexistence
switches into a coexistence of  SC and SDW phases, due to M-SC phase transition.   A gradient in shading (SDW/SC region) below $T_{\text {SC}}$ denotes the increase in volume proportion of SC phase in the  bulk sample.  In inset:  Our diagram is completed with data taken from  Bi\v{s}kup {\it et al.}~(open points). Solid curve is the fit to our empirical formula for $T_{\text {SDW}}$ \vs $p$ dependence (Eq.~\ref{TvspBest}).}
\label{Plot7}
\end{figure}

Our data combined with the study of Bi\v{s}kup~\al which 
have determined normal state-to-SDW transition temperatures 
in the range 1 bar to 7.5 kbar, allow to present a $p,T$ 
phase diagram of \pff~displayed in Fig.~\ref{Plot7}. The fit  
to an empirical formula which  takes into account the fact 
that $p_c$ is found at $T_{\text {SC}} =1.2$~K, and not at 
$T=0$ K leads to

\begin {equation} T_{\text {SDW}}(p)=T_1-[(T_1-T_{\text 
{SC}})*(p/p_c)^3] 
\label{TvspBest} 
\end {equation}

Here $T_{\text {SC}}$ is the experimental value whereas 
$T_1= T_{\text {SDW}}(1 bar)$ and $p_c$ are free parameters. 
Best parameter values are $T_1=12.0\pm0.15$~K and 
$p_c=9.43\pm0.04$~kbar. Obviously, $T_1$ and $p_c$ values 
correspond excellently to the experimental ones despite the 
fact that they were obtained as the only  free parameters in 
the fit. It is also interesting to note that the $T_{\text 
{SDW}}$ pressure dependence seems to be a pure cubic one.

\subsection{Quantification of the domain fraction in the inhomogeneous SDW regime} 
There are several features in our experimental data substantiating the coexistence of phases, SDW/M above 
1.2 K and SDW/SC below 1.2 K. Hysteresis presented in the Fig.~\ref{Plot4} is the primary, still only qualitative one. But the other two features (besides evidencing for the coexistence) lead us to the possibility of quantification of the phases volume fraction in the bulk. Indeed, we consider the orders of magnitude change in the critical current, as presented in Figs.~\ref{Plot5} and~\ref{Plot6} directly proportional to the change in the effective cross-section taken up by superconducting domains for SDW/SC coexistence situation. For the SDW/M situation we intend to show that the change in the effective cross-section taken up by
metallic domains may be incurred from the orders of magnitude change in the resistance of the
sample for pressures inside the 8.65 -- 9.43 kbar region.

In the following we make the assumption  that the sample in the inhomogeneous regime behaves
as a composite made of two materials. The two materials having the properties of
the sample at 11 kbar (pure SC ground state) and 6.8 kbar (pure SDW ground state) respectively.

\subsubsection{ $T<T_{\text {SC}}$: Quantification of the SDW/SC domain fraction}

 At 11 kbar, the pressure is sufficiently far above the critical pressure to have a fully
homogeneous state, either metallic or SC. Measurements of the critical current from 0.4 to 1.0
K, at this pressure, emphasize this (Fig.~\ref{Plot6}b). That is, one may divide U/I for the
highest currents, which completely suppress SC state, and will recover a resistance which is
comparable to the resistances measured above $T_{\text {SC}}$ at this same pressure (values of
order of m$\Omega$). Obviously, high currents take the sample from purely superconducting state
to a purely metallic state. We conclude then that superconductivity is here homogeneous through
the whole cross-section of the sample with a critical current density of $J_{c}=200$~A/cm$^{2}$.
Since $T_{\text {SC}}$ is nearly constant in the whole pressure range of the inhomogeneous regime, so
should the critical current density. 

In the inhomogeneous regime, we can model the sample as alternating insulating (SDW)  channels
and free electrons (eventually superconducting) channels. For simplicity, the channels are
assumed to extend longitudinally from one end of the sample to the other. A change of pressure
is equivalent to a change of the  cross-section of the metallic (or SC) channels and of the SDW
channels. We use the crude approximation  that c is temperature independent. In the
inhomogeneous regime, in the absence of any weak links along the conducting channels, the SC
fraction of the sample cross-section is  given by the respective $I_c$ value divided by 
$I_{cMAX}$ measured at 11 kbar, in the pure SC state. Accordingly, at pressures in the
inhomogeneous regime, the critical current is lowered, \eg~at 9.1 kbar, $I_c=7$~mA
(Fig.~\ref{Plot6}a). The respective values for other pressures are given  in the Table 1.

The resistance recovered for  currents above the critical value at 9.1 kbar is only 12
m$\Omega$, \ie~about 25 times smaller than the resistance of the sample just above
$T_{\text {SC}}$ at this same pressure (see Fig.~\ref{Plot3}). This feature can be ascribed to
an  electric field overcoming the field required for the  depinning of the SDW state. Actually,
we calculate the electric field to be of the order of 20 mV/cm at a measuring current of 7 mA
($T=$1 K) using the data for the 9.1 kbar  pressure run. This electric field is about four
times the  value of the depinning field measured in \pff~\cite{tomic91}  or (TMTSF)$_2$AsF$_6$,
\cite{traetteberg94} at ambient  pressure. The conductance at high currents is therefore the 
sum of the sliding SDW conductance (which still depends  weakly on current) and the conductance
of the {\em  decondensated} free electrons in the suppressed SC, now metallic domains. 

\begin{figure}
\centering\resizebox{0.46\textwidth}{!}{\includegraphics*{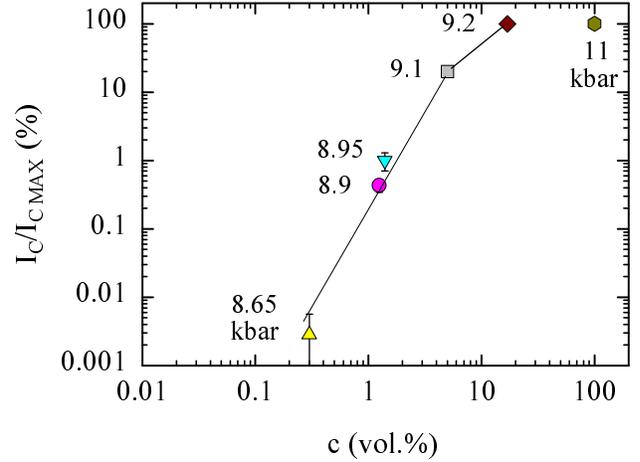}}
\caption{Correlation between the critical current $I_c$ (at 0.4 K) and the volume fraction $c$~(vol.\%) of the metallic phase in the bulk sample  which is present between $T_{\text {SC}}$ and  $T_{\text {SDW}}$ in the coexistence regime. Values of both variables are given in Table~\ref{tablica1}.}
\label{Plot8}
\end{figure}

\subsubsection{ $T_{\text {SC}}<T<T_{\text {SDW}}$: Quantification of the SDW/M domain fraction}

The principle of additive conductances leads us to a second, independent procedure, for
quantification of the SDW/non-SDW domain fraction. We come back to the resistance curves and
extract  the non-SDW fraction, c, and SDW fraction, 1-c, from the resistance data in the
$T_{\text {SC}}<T<T_{\text {SDW}}$ temperature range. Again, we model the sample as alternating
insulating (SDW)  channels and free electrons (eventually superconducting) channels. Thus, the
fraction parameter c is related to effective cross-sections, not the volume of domains. Using
this "rigid model",  the Arrhenius law conductivity can be recalculated as:

\begin{equation}
\sigma(T)=c\sigma_m+(1-c)\sigma_{\text SDW}
\label{sigmaT}
\end{equation}

In terms of the resistance we get:
\begin{equation} 1/R(T)=c/R_m+(1-c)/R_{\text SDW}
\label{rhoT}
\end{equation}

where $R_{m}$ is the resistance of a 100\%  metallic sample 
and $R_{\text {SDW}} = R_{\infty}exp(\Delta /T)$. Fitting 
the experimental data  $\log R$ \vs $1/T$ to Eq.~\ref{rhoT} 
gave quite good fits as presented in Fig.~\ref{Plot3} with 
the series of dotted lines. The fit parameters c, $\Delta $ 
and $R_{\infty }$ are given in the Table~\ref{tablica1}. The 
observed decrease of $R_{\infty }$ with pressure is the 
clear demonstration of the increase of the metallic fraction of the sample.
The observed 
evolution of $\Delta$ with pressure is less clear. In pure 
SDW regime it decreases with pressure, scaled with 
the transition temperature. Indeed $2\Delta=2.4T_{\text 
{SDW}}$, although the BCS factor would be 3.52. Further, in 
the inhomogeneous SDW regime $\Delta$ assumes more or less 
constant value. We attribute such a result to the fact that the 
fitting procedure was performed in rather narrow 
temperature spans, $T_{\text {SC}}<T<T_{\text  {SDW}}$.

\subsubsection{ Correlation of the two quantification procedures}
 
Our study shows that the coexistence regime extends over the 
8.65 kbar--9.43 kbar range ($\Delta p=0.78$ kbar). 
Fig.~\ref{Plot8} shows the correlation between  the two 
independent quantifications of the non-SDW fraction c. These 
were obtained either by critical current measurements in the 
SC state (y-axis, $I_{c MAX}=$35 mA), or by resistance 
measurements (x-axis, c from the recalculated Arrhenius law) for $T_{\text {SC}}<T<T_{\text {SDW}}$ range. We therefore 
plausibly demonstrate the existence of two regimes: a higher 
inhomogeneous regime (9.1--9.43 kbar) where SC domains 
extend from one end of the sample to the other, and a lower 
inhomogeneous regime (8.65--9.1 kbar) where Josephson 
junctions (or phase slip centers \cite{Tinkham}) are present 
along  the conducting channels and further reduce the 
critical current to lower values.

\subsection{The phase segregation scenario}

The problem of the metal-insulator transition under pressure has been extensively
studied in the context of the canonical Mott transition \cite{Mott}. The picture
proposed by Mott relies on the existence of a first order phase transition
between  homogenous metallic and insulating phases. According to Mott there
exists a  critical pressure at which the specific volume jumps discontinuously 
 from its value in the insulating state to the value in the metallic one. The
model proposed by Mott did not require a phase segregation regime but only a
volume discontinuity from one homogenous state to the other. The situation for
\pff is quite different as experimental data presented above show the existence
of a rather broad pressure domain in which the two different orders coexist
therefore ruling out the picture of the canonical Mott model. Hence it is the
experimental data which have imposed the search of an other  approach based on
a variational theory leading to an inhomogeneous phase with an energy lower than
the energy of homogenous states (SDW or Metal). In the following, we propose a
discussion of the  experimental data in the framework of the Fermi liquid 
picture, which is expected to be valid in this low  temperature part of the phase
diagram. We have to study the  relative stabilities of different phases: the
metal, the  spin density wave, the superconducting phase, but also, of  course,
possible phases where coexist either a SDW phase and  a metallic phase, or, at
lower temperature, a SDW phase and  a superconducting one.

In order to discuss the stability of the SDW phase, it is therefore essential
to  model  the Fermi surface geometry, \ie  the non
interacting electrons dispersion relation.\ We shall describe the non
interacting electron gas by the following two-dimensional dispersion
relation :
\begin{equation}
\varepsilon \left( \mathbf{k}\right) =v_{F}\left( \left| k_{x}\right|
-k_{F}\right) +t_{\bot }\left( k_{y}b\right)
\end{equation}
This dispersion has been linearised around the Fermi level, along the
direction $\mathbf{a}$ , corresponding to the highest conductivity, $v_{F}$
is the Fermi velocity and $t_{\bot }\left( k_{y}b\right) $ describes the
periodic warping of the Fermi surface along $\mathbf{b}$ the second
direction of high conductivity. The third direction, which does not play a
major role in this problem, has been ignored.\ In general, $t_{\bot }\left(
k_{y}b\right) $ is responsible for a nearly perfect Fermi surface nesting,
which means that there exist a nesting wave vector $\mathbf{Q}$ obeying :
\[
\varepsilon \left( \mathbf{k}\right) \simeq -\varepsilon \left( \mathbf{k}+%
\mathbf{Q}\right)
\]
for any wave vector $\mathbf{k}$ of the Fermi surface. The simplest model is
given by the following choice \cite{GorkovLebed}\cite{Montambaux2} :

\begin{equation}
t_{\bot }\left( p\right) =-2t_{b}\cos p-2t_{b}^{^{\prime }}\cos 2p
\end{equation}
For $t_{b}^{^{\prime }}=0,$ the Fermi surface is perfectly nested, with wave
vector $\mathbf{Q}_{1}=\left( 2k_{F},\pi /b\right) .\;$Deviations from
perfect nesting, which destabilize the SDW phase, are described by a
unique parameter $t_{b}^{^{\prime }}$.\ This is an oversimplified
description of the Fermi surface geometry, which can be improved by taking
into account multiple site transverse transfer integrals\cite{Yamaji1}.
However, we limit, here, our discussion to this simple one-parameter model,
which gives the essential physical features.\ The straightforward extension
to a multiple transverse transfer integral model will be given in a
forthcoming paper. The critical temperature for the formation of the SDW
phase, $T_{\text {SDW}}$ is given by
\begin{equation}
\chi ^{0}\left( \mathbf{Q},t_{b}^{^{\prime }},T_{\text SDW}\right) =1/\lambda
\label{chi0}
\end{equation}
where $\chi ^{0}$ is the susceptibility of the non interacting electrons and
$\lambda $ is a phenomenological parameter describing the strength of the
electron interaction. The wave vector of the magnetic order $\mathbf{Q=}%
\left( Q_{x},Q_{y}\right) $ should correspond to the maximum of $\chi
^{0}\left( \mathbf{Q},t_{b}^{^{\prime }},T_{c}\right) .\;$In our simple
model, we obtain the well known result :
\begin{eqnarray}
\chi ^{0}\left( \mathbf{Q},t_{b}^{^{\prime }},T\right) & = & N\left( 0\right) \times
\nonumber\\ & \times & \left[ \ln \frac{E_{0}}{T}+\psi (\frac12)-Re\left\langle \psi(\frac12+iB) \right\rangle \right]
\end{eqnarray}
where $B=A/4\pi T,N\left( 0\right) =1/\left( 2\pi v_{F}b\right) $ is the
density of states at the Fermi level, \ $\left\langle \cdots \right\rangle $
means averaging over the transverse momentum $p$, $E_{0}$ is a cutoff
proportional to the bandwidth, $\psi $ is the digamma function and :
\begin{equation}
A\left( p\right) =Q_{x}-2k_{F}+\left( 1/v_{F}\right) \left[ t_{\bot }\left(
p\right) +t_{\bot }\left( p-Q_{y}b\right) \right]
\end{equation}

Several authors\cite{Yamaji2}\cite{Hasegawa} have considered two different
SDW wave vectors.\ At zero temperature, the best nesting wave vector,
denoted $\mathbf{Q}_{2}$ in the literature, connects the inflexion points of
the Fermi surface.\ It is given by :
\begin{equation}
A\left( p,\mathbf{Q}_{2}\right) =\frac{\partial }{\partial p}A\left( p,%
\mathbf{Q}_{2}\right) =\frac{\partial ^{2}}{\partial p^{2}}A\left( p,\mathbf{%
Q}_{2}\right) =0
\end{equation}
At high temperature, when $T$ becomes of the order of $t_{b}^{^{\prime }}$ ,
deviation from perfect nesting becomes irrelevant and the best nesting wave
vector is $\mathbf{Q}_{1}=\left( 2k_{F},\pi /b\right) .\;$According to
Hasegawa and Fukuyama\cite{Hasegawa}, $\mathbf{Q}_{2}\left( T\right) $
varies with temperature and jumps to $\mathbf{Q}_{1}$ at a critical
temperature $T^{*}=0.232t_{b}^{^{\prime }}$ (for $t_{b}/t_{b}^{^{\prime }}=20%
\sqrt{2}$ ). However, as far as we know, no clear experimental evidence of
this first order transition has been given.

We first discuss the stability of the \textit{homogeneous} SDW$_{1}$ phase
with non varying wave vector $\mathbf{Q}_{1}$.\ When applying a pressure,
the essential feature for this stability is the increase of $t_{b}^{^{\prime
}}.$ The critical line for the homogeneous order is given by the Stoner
criterion (Eq. \ref{chi0}).\ For a given interaction $\lambda ,$ there is a
critical
line value $t_{b}^{^{\prime }*}\left( T\right) ,$ which means a critical
pressure $p_{c}\left( T\right) $ at which the magnetic order disappears. The
ratio $T_{SDW}/t_{b}^{^{\prime }*},$ where $T_{SDW}$ is the magnetic
critical temperature, is a universal function of $\beta =t_{b}^{^{\prime
}}/t_{b}^{^{\prime }*}$ \cite{Yamaji2}\cite{Hasegawa}.

 What is revealed, in fact, by the experimental data described above, is
the simultaneous formation of two different phases SDW/Metal or SDW/SC,
according to the temperature.\ However, this
coexistence corresponds to a segregation in the direct space, not in the
reciprocal space. It is quite plausible that such a segregation is not
produced on a microscopic scale ($l\ll \xi $ , where $\xi $ is the
correlation length) or a mesoscopic scale( $l\sim \xi ),$ but rather on a
macroscopic scale ( $l\gg \xi $ ), which is much more favorable to the
carrier localisation energy necessary to spatially confine the electrons,
but also to the interface energy necessary to create domain walls between
regions of different orders.

We give very simple and general arguments proving that, near enough to the
critical line for the formation of an homogeneous SDW phase, a \textit{%
spatially heterogeneous} phase has a lower free energy than the \textit{%
homogeneous} SDW phase.\ The origin of such a phenomenon is due to the
following essential physical features :
\begin{itemize}
\item The relevant quantity on which depends the SDW order stability is 
$t_{b}^{^{\prime }}$. Applying a pressure increases $t_{b}^{^{\prime }}$
and, therefore, the SDW free energy $F_{m}\left( t_{b}^{^{\prime }}\right)$, 
up to a critical value $t_{b}^{^{\prime }*}$ at which the homogeneous SDW
phase disappears.

\item The SDW stability decreases very strongly near $t_{b}^{^{\prime }*}.\;$%
The slope of the critical line is very large.

\item The relevant quantity to stabilize the SDW phase near $t_{b}^{^{\prime
}*}$ is $b,$ the unit cell parameter along the $y-$direction$.\;$Indeed,
increasing $b$ strongly decreases $t_{b}^{^{\prime }},$ and, therefore,
strongly lowers $F_{m}\left( t_{b}^{^{\prime }}\right) .$

\item It is always favorable to create a heterogeneous phase : one part, the
volume of which is $\left( 1-c\right) \Omega ,$ has a cell parameter $%
b+\delta b_{1}$ and is magnetic, with a lower magnetic free energy because
of the lower $b$ parameter ; the other part, the volume of which is $c\Omega$, 
is metallic and has a cell parameter $b-\delta b_{2}$,($\Omega$ is the
total volume), which imposes $\delta b_{2}/\delta b_{1}=\left( 1-c\right)
/c$. The latter relation implies the constant volume assumption which is
considered in the present model (\textit{see also Sec.5 for additional
arguments)}.  The elastic energy cost for such a deformation is, indeed,
proportional to $\left(
\delta b\right) ^{2}$ and, therefore of second order, while the deformation
allows to gain a magnetic free energy proportional to the first order quantity
$\left(
\frac{\partial F_{m}}{%
\partial t_{b}^{^{\prime }}}\right) \left( \frac{\partial t_{b}^{^{\prime }}%
}{\partial b}\right) \delta b.\;$The larger the slope$\left( \frac{\partial %
F_{m}}{\partial t_{b}^{^{\prime }}}\right) ,$ the larger the free energy
lowering.
\end{itemize}

The physical picture is clear : near enough to the ``homogeneous critical
line'', for a pressure lower than the critical pressure, the formation of
macroscopic metallic domains or ``ribbons''(in this 2D model), with a lower $%
b$ parameter, parallel to the direction of highest conductivity, allows the
formation of SDW ``ribbons'', which on the contrary have a larger $b$
parameter.\ The elastic energy cost is
\begin{eqnarray}
\Delta E_{elastic} &=&\left( 1-c\right) K\left( \delta b_{1}\right)
^{2}+cK\left( \delta b_{2}\right) ^{2}  \nonumber \\
&=&\frac{1-c}{c}K\left( \delta b_{1}\right) ^{2}
\end{eqnarray}
where $K$ is an elastic constant.\ The magnetic free energy lowering,
compared to the homogeneous phase free energy, is given by
\begin{equation}
\Delta F_{m}=\left( 1-c\right) \left( \frac{\partial F_{m}}{\partial %
t_{b}^{^{\prime }}}\right) \left( \frac{\partial t_{b}^{^{\prime }}}{%
\partial b}\right) \delta b_{1}-cF_{m}\left( t_{b}^{^{\prime }}\right)
\end{equation}
This linear approximation is quite satisfactory because the derivative is
quite large. Minimizing the total free energy $\Delta F_{total}=\Delta %
E_{elastic}+\Delta F_{m}$ with respect to $\delta b_{1}$ and c , we find
that the stable phase is \textit{heterogeneous}, with a fraction c of
metallic phase forming macroscopic metallic domains parallel to $\mathbf{a}$
:
\begin{equation}
c=\frac{1}{2}-\frac{2K\;sgn \left(  t_{b}^{^{\prime }}-t_{b}^{^{\prime
}*}\right) \left| F_{m}\left( t_{b}^{^{\prime }}\right) \right| }{%
\left( \frac{\partial F_{m}}{\partial t_{b}^{^{\prime }}}\right) ^{2}\left(
\frac{\partial t_{b}^{^{\prime }}}{\partial b}\right) ^{2}}
\end{equation}

The free energy of the heterogeneous phase is given by
\begin{equation}
\Delta F_{total}=-\frac{1}{4}\frac{\left[ \frac{1}{4K}\left( \frac{\partial %
F_{m}}{\partial t_{b}^{^{\prime }}}\right) ^{2}\left( \frac{\partial %
t_{b}^{^{\prime }}}{\partial b}\right) ^{2}-F_{m}\left( t_{b}^{^{\prime
}}\right) \right] ^{2}}{\frac{1}{4K}\left( \frac{\partial F_{m}}{\partial %
t_{b}^{^{\prime }}}\right) ^{2}\left( \frac{\partial t_{b}^{^{\prime }}}{%
\partial b}\right) ^{2}}<0
\end{equation}

On the ``homogeneous critical line'', the fraction of metallic phase is $%
c=1/2.$ It decreases as $t_{b}^{^{\prime }}$ or the applied pressure
decreases and vanishes when :
\begin{equation}
\left| F_{m}\left( t_{b}^{^{\prime }}=t_{b1}^{^{\prime }}\right) \right| =%
\frac{1}{4K}\left( \frac{\partial F_{m}}{\partial t_{b}^{^{\prime }}}\right)
^{2}\left( \frac{\partial t_{b}^{^{\prime }}}{\partial b}\right) ^{2}
\end{equation}
which defines $t_{b1}^{^{\prime }}$ , the lower critical pressure for the
formation of the heterogeneous phase. For $t_{b}^{^{\prime
}}<t_{b1}^{^{\prime }}$ , the stable phase is an homogeneous SDW phase.

In a symmetrical way, for $t_{b}^{^{\prime }}$ larger than the critical
value, i.e. when the homogeneous phase is metallic, the total free energy is
lowered by the formation of SDW macroscopic domains, forming magnetic
``ribbons'' along the $\mathbf{a}$ direction. An expression quite similar to
the case $t_{b}^{^{\prime }}<$ $t_{b}^{^{\prime }*}$ can be obtained.\ We
find again that $c=1/2$ on the ``homogeneous critical line'' and increases
with $t_{b}^{^{\prime }},$ \ie with the applied pressure and goes to $1$
when :
\begin{equation}
F_{m}\left( t_{b}^{^{\prime }}=t_{b2}^{^{\prime }}\right) =\frac{1}{4K}%
\left( \frac{\partial F_{m}}{\partial t_{b}^{^{\prime }}}\right) ^{2}\left(
\frac{\partial t_{b}^{^{\prime }}}{\partial b}\right) ^{2}
\end{equation}
which defines $t_{b2}^{^{\prime }}$ , the upper critical pressure for the
formation of an heterogeneous phase.\ For $t_{b}^{^{\prime
}}>t_{b2}^{^{\prime }}$ , the stable phase is an homogeneous metal.

\begin{figure}
\centering\resizebox{0.48\textwidth}{!}{\includegraphics*{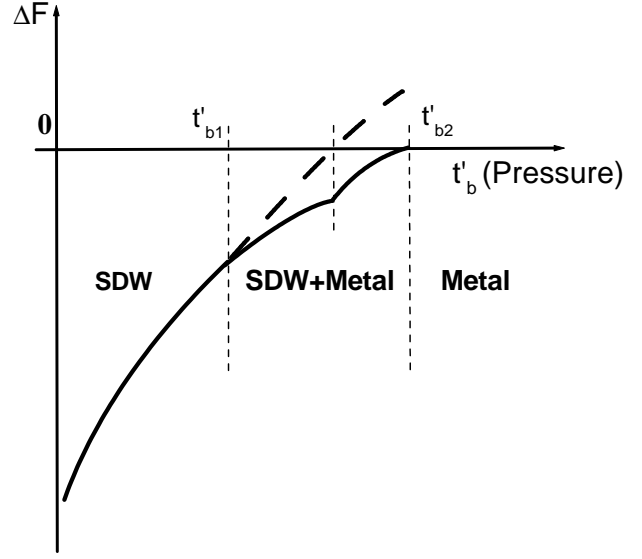}}
\caption{Schematic behaviour of the energy in the instability domain. The x-axis is
$t_{b}^{^{\prime }}$ but can be the applied pressure as well.}
\label{Plot9}
\end{figure}

In Fig.~\ref{Plot9} we have displayed schematically the pressure dependence of the magnetic
condensation energy and the stability domain of the spatially modulated phase with the energy
of the inhomogenous phase (continuous line) which is lower than that of the unstable homogenous
phase (dashed line).

It should be stressed that we did not take into account the energy necessary
to form domain walls between the metallic and magnetic domains.\ We did not
take into account either the energy necessary to localize the carriers
within a domain.\ Since the domains are believed to be macroscopic, the
corresponding corrections should be quite small.

Clearly, at lower temperatures, the macroscopic metallic domains should
undergoe a transition to a superconducting order, since the phase diagram
exhibits a competition between magnetic and superconducting orders.\ We
have, therefore, to investigate the formation of an heterogeneous phase,
with coexisting SDW and superconducting domains. Similar calculations can
be done by including the superconducting phase free energy, $F_{s},$ which
will be assumed to be independent of $t_{b}^{^{\prime }}.$ Such an
approximation is certainly valid in the narrow pressure range considered
here. We shall not make any assumption, neither on the physical mechanism
inducing the superconductivity, nor on the symmetry of the order parameter.
We shall only suppose that $F_{s}$ is given by the usual mean field
expression. As above, we find that the total free energy is lowered by the
formation of an heterogeneous phase, on both sides of the ``homogeneous
critical line'', with a volume $c$ of superconducting domains and a volume $%
\left( 1-c\right) $ of SDW domains. $c$ vanishes for $t_{b}^{^{\prime }}%
\leq $ $t_{b3,}^{^{\prime }}$ where the lower critical value $%
t_{b3}^{^{\prime }}$ is given by :
\begin{equation}
-F_{m}\left( t_{b3}^{^{\prime }}\right) =\frac{1}{4K}\left( \frac{\partial %
F_{m}}{\partial t_{b}^{^{\prime }}}\right) ^{2}\left( \frac{\partial %
t_{b}^{^{\prime }}}{\partial b}\right) ^{2}-F_{s}
\end{equation}
Since $F_{s}<0,$ the coexistence region gets broader when the
superconducting order grows as $T\rightarrow 0.\;$In the same way, in the
superconducting region of the homogeneous phase diagram, the total free
energy is lowered by the formation of a fraction$\left( 1-c\right) $ of SDW
domains. On the ``homogeneous critical line'', $c=1/2$ and increases when $%
t_{b}^{^{\prime }}$ increases, up to an upper critical value $%
t_{b4}^{^{\prime }}$ , given by ;
\begin{equation}
F_{m}\left( t_{b4}^{^{\prime }}\right) =\frac{1}{4K}\left( \frac{\partial %
F_{m}}{\partial t_{b}^{^{\prime }}}\right) ^{2}\left( \frac{\partial %
t_{b}^{^{\prime }}}{\partial b}\right) ^{2}+F_{s}
\end{equation}

In order to calculate explicitly the critical lines, it is necessary to
evaluate the magnetic free energy $F_{m}\left( t_{b}^{^{\prime }}\right) .$
Its expression can be found in the literature.\ Near the critical line, a
Landau expansion can be written\cite{Montamb40} :
\begin{equation}
F_{m}=\frac{8\pi T_{SDW}^{2}N\left( 0\right) }{Re\left\langle \psi
"\left( \frac{1}{2}+iB\right) \right\rangle }\left[ \frac{\partial \chi ^{0}%
}{\partial T}dT+\frac{\partial \chi ^{0}}{\partial t_{b}^{^{\prime }}}%
dt_{b}^{^{\prime }}\right]
\label{landau}
\end{equation}
The quantities $t_{b}^{^{\prime }*}\frac{\partial \chi ^{0}}{\partial T}$
and $t_{b}^{^{\prime }*}\frac{\partial \chi ^{0}}{\partial t_{b}^{^{\prime }}%
}$ are well known universal functions of $\beta =t_{b}^{^{\prime
}}/t_{b}^{^{\prime }*},$ given in the literature\cite{Montamb40},
\begin{eqnarray}
t_{b}^{^{\prime }*}\frac{\partial \chi ^{0}}{\partial T} &=&-N\left( 0\right) \frac{t_{b}^{^{\prime }*}}{T_{SDW}}
\times  \nonumber\\ & \times & 
\left[ 1+Im\left\langle
B\psi ^{\prime }\left( \frac12+iB\right) \right\rangle \right]  \\
t_{b}^{^{\prime }*}\frac{\partial \chi ^{0}}{\partial t_{b}^{^{\prime }*}}
&=&-N\left( 0\right) \frac{t_{b}^{^{\prime }*}}{T_{SDW}} \times  \nonumber\\ & \times & Im%
\left\langle \pi^{-1}\cos \left( 2p\right) \psi ^{\prime } \left( \frac12%
+iB\right) \right\rangle
\end{eqnarray}
where $\psi ^{\prime }$ is the trigamma function.\ These expressions allow a
complete determination of $F_{m}$ close to the critical line.

\begin{figure}
\centering\resizebox{0.44\textwidth}{!}{\includegraphics*{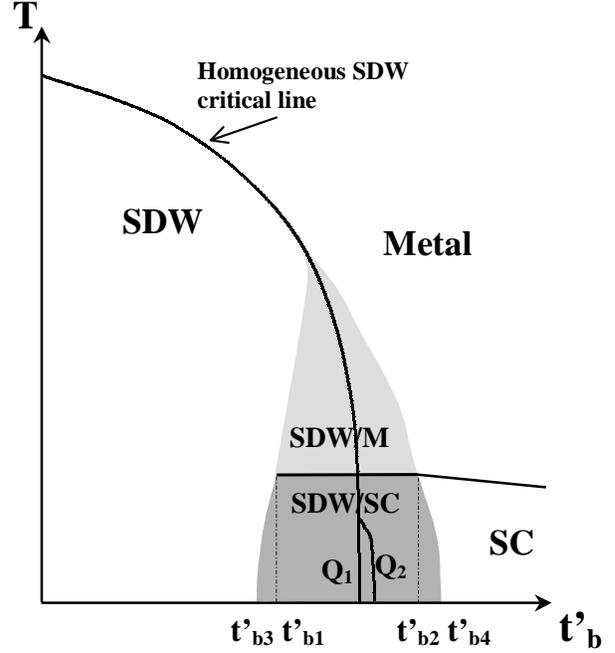}}
\caption{Schematic phase diagram showing the inhomogeneous SDW/M and SDW/SC phases in the
vicinity of the critical border. The x-axis can be the applied pressure as well. The heavy
lines in the homogenous phase region correspond to $\mathbf{Q}_{1}$ and $\mathbf{Q}_{2}$ SDW
states.}
\label{Plot10}
\end{figure}

In the limit of vanishing temperature, the magnetic free energy becomes the SDW
energy :
\begin{equation}
F_{m}\left( t_{b}^{^{\prime }}\right) \rightarrow N\left( 0\right) \left[
\left( t_{b}^{^{\prime }}\right) ^{2}-\left( M_{0}\right) ^{2}\right]
\label{t0}
\end{equation}
where $M_{0\text{ }}$is the magnetic order parameter at zero temperature.
Expressions (\ref{landau}-\ref{t0}), together with the standard mean field
expression for $F_{s}$ allow a complete determination of the critical lines 
$c=0$ and $c=1$, 
which give the limits of stability of the heterogeneous phase in the 
$(T,t_{b}^{^{\prime }})$ plane, or, equivalently in the $(T,p)$ plane. A schematic
illustration of the theoretical phase diagram is displayed on Fig.~\ref{Plot10}.

We would like to emphasise that, within our model, we do not expect any
pressure variation of the superconducting critical temperature $T_{c}$, as
experimentally observed, in the heterogeneous phase. It is possible to
interpret $T_{c}$ as the ordering temperature of the metallic domains.
Nevertheless, we expect the superconducting ordering increases the width of
the pressure range in which the heterogeneous phase is stable, because it
lowers the total free energy.

The maximum width of the stability region of the heterogeneous phase is
obtained at zero temperature.\ We consider the following numerical values of
the parameters entering the model : $E_{0}=3000K,$ $t_{b}^{^{\prime }}=10K$
\cite{Montambaux2}\cite{Montambaux3}, $\frac{\partial t_{b}^{^{\prime }}}{%
\partial p}=1$K/kbar \cite{Grant}\cite{Ducasse}, $\frac{1}{b}%
\frac{\partial b}{\partial p}=3$ $10^{-3}$ kbar$^{-1}$\cite{these}.
These values can be considered as typical. We obtain a pressure range of
stability of the heterogeneous phase of the order of 1 kbar. The
agreement with experiments can be considered as extremely good, if we
consider the crudeness of the model.

We discuss now the case of the SDW$_{2}$ phase corresponding to the wave
vector $\mathbf{Q}_{2}$ Near to the transition between SDW$_{1}$ and SDW$_{2}
$ studied by Hasegawa and Fukuyama\cite{Hasegawa}, a Landau expansion of the
magnetic free energy similar to equation (\ref{landau}) can be given, but with
different values of $\frac{\partial \chi ^{0}}{\partial T}$ and $\frac{%
\partial \chi ^{0}}{\partial t_{b}^{^{\prime }}}.$ It is known that the
absolute value of the critical line slope is smaller for the SDW$_{2}$ phase%
\cite{Montamb40} \cite{Hasegawa}.\ For that reason, the free energy lowering
associated to the formation of an heterogeneous phase is larger when the
magnetic phase is the SDW$_{1}$ phase.\ We therefore expect that the
formation of the heterogeneous phase SDW$_{1}+$ Metal precludes the
formation of the SDW$_{2}$ phase, at least near the region of transition
between SDW$_{1}$ and SDW$_{2}$

It is also possible to discuss the stability of the SDW$_{2}$ phase in the
limit of zero temperature. Hasegawa and Fukuyama\cite{Hasegawa}, as well
as Yamaji\cite{Yamaji2}, have studied theoretically the stability of an
homogeneous SDW phase. As a result, they predict the stability of an
homogeneous SDW$_{2}$ phase in a narrow pressure range, for $p$ larger than
the critical value $p_{c1}$ at which SDW$_{1}$ disappears, but below a
second critical pressure $p_{c2.}$ However, in this pressure range, while
the magnetic free energy $F_{m}\left( t_{b}^{\prime }\right) $ is slightly
lower for the SDW$_{2}$ phase, the absolute value of the slope $\left| \frac{%
\partial F_{m}}{\partial t_{b}^{^{\prime }}}\right| $ is much smaller than in
the SDW$_{1}$ phase and almost vanishes at $p=p_{c2}.$ The lower slope leads
to an energy decrease associated to an heterogeneous phase which is lower
than in the case of the SDW$_{1}$ phase and vanishes as $p\rightarrow p_{c2}$. 
Using the same model of non interacting Fermi surface as ours, Hasegawa
and Fukuyama calculated that the SDW$_{2}$ phase is stable up to $%
t_{b}^{^{\prime }}=1.03t_{b}^{^{\prime }*}$ \cite{Hasegawa}, which leads to
a pressure range for the SDW$_{2}$ stability certainly smaller than 0.3
kbar.\ Yamaji obtained a similar result for the same Fermi surface\cite
{Yamaji4}. If this result is correct, within this pressure range, the
heterogeneous phase SDW$_{1}+$ SC phase ($Het_{1}$ phase) is
quite probably more stable than the homogeneous SDW$_{2} (Hom_{2}$ phase).
In such a case, the latter magnetic phase could not be observed, which might
explain why no experimental signature of a transition between different
magnetic phases has been seen in this part of the phase diagram. However, it
should be stressed that Yamaji has studied the stability of the SDW$_{2\text{
}}$phase using a more realistic Fermi surface model, which seems more
favourable to the SDW$_{2}$ stability\cite{Yamaji2}.\ Further work is
necessary to determine the relative stability of $Het_{1}$ and $Hom_{2}$
phases, when a realistic Fermi surface is taken into account.

\section{Conclusion}

In summary, this new visit to the $p,T$ phase diagram of 
\pff~enables us to reach a better understanding of the 
cross-over between SDW and SC ground states. The 
experimental results suggest that  a picture of coexisting 
SDW and SC macroscopic domains prevails in a narrow pressure 
domain of $\approx 0.8$ kbar below the critical pressure 
marking the establishment of an homogeneous SC ground state. 
In spite of a volume fraction of a SC phase strongly 
depressed at decreasing pressure in the coexistence regime 
we could not detect any significant change of $T_{\text {SC}}$.The 
early claim for the absence of SDW/SC coexistence in the 
vicinity of the critical pressure \cite{azevedo84} based on 
the observation of an EPR response typical for the 
superconducting instability, can now be understood by the 
impossibility of the EPR technique to observe the very 
broad signal coming from the SDW domains.

We have proposed a very simple model which, on the basis of 
quite general arguments, leads to the formation of a 
heterogeneous phase, near the critical line of the 
homogeneous magnetic phase, on both sides of it, where 
coexist metallic and magnetic domains and, at lower $T$, 
magnetic and superconducting domains. This result provides a 
quite plausible interpretation of the data reported here. 
Obviously, the same kind of arguments might apply to other 
competing instabilities. What we are  proposing is a variational theory which says that it is
possible to find  an inhomogenous phase with a free energy  which is lower than the energy of
the homogenous states (SDW or metal). We took for simplicity the assumption  that the total 
volume of the sample is kept constant. In any variational calculation  some constraints are,
of  course, necessarily imposed on the trial wave functions or trial states. Here, we have
choosen, as a variational constraint, the constraint of a constant total volume, for the sake
of simplicity. Our variational calculation shows that a state exhibiting a "coexistence"  (or
phase segragation) has a lower free energy than the homogenous phase. However, such a free
energy lowering is not due to the constraint. If we relax the constraint, the free energy of
the inhomogenous state could still decrease and be even more stable than that of the homogenous
state. But then, the model should rely very much on the detailed pressure dependence of the  SDW
condensation energy versus pressure. 

As far as organics are concerned, it is interesting to 
mention that similar phenomena have been reported to 
arise in the recently discovered superconductor of the 
TM$_{2}$X family, (TMTTF)$_2$PF$_6$, under very high 
pressure \cite{wilhelm01} but the extreme pressure 
conditions of the latter compound made a detailed study 
impossible. An other situation may be encountered with the 
coexistence of two possible anion orderings namely 
(1/2,1/2,1/2) and (0,1/2,1/2) in the salt (TMTSF)$_2$ReO$_4$ 
under pressure detected by x-ray scattering \cite{moret86}. 
We may anticipate that the coexistence domain will also be 
characterised by a segregation between metallic regions 
(associated to the (0,1/2,1/2) order) and (1/2,1/2,1/2) 
anion-ordered insulating regions \cite{tomic89}. A 
coexistence between SDW and SC orders has also been 
mentioned in this latter compound in the narrow pressure 
domain $8\pm 0.25$ kbar \cite{tomic89}.

Furthermore, an immediate consequence
of the existence of magnetic macroscopic domains in the superconducting
phase is a large increase of the upper critical field $H_{c2}$ which had been
reported
long time ago by Greene and Engler in \pff \cite{greene80} and by Brusetti \al
 in \Aff \cite{brus82}.
Finally, Duprat and Bourbonnais \cite{duprat01} have recently performed a
calculation of the  interference between
SDW and SC channels using the formalism of the renormalisation group at
temperatures below the 1D cross-over in the
presence of strong deviations to the perfect nesting. This calculation shows the
possibility of reentrant
superconductivity in the neighbourhood of the critical pressure and can explain
the deviations of the $\Delta /T_{\text {SDW}}$
ratio from the BCS ratio in the vicinity of the critical pressure due to the
non-uniform character of the SDW gap over
the Fermi surface (although similar deviations are also obtained, for the same physical reason,
in a simple mean field treatment \cite{gorkov83}). Very likely, merging the results of
reference
\cite{duprat01} with the  present coexistence model could improve the overall
theoretical description.

\begin{acknowledgement}
We would like to thank very warmely M. Ko\'{n}czykowski for providing the
sensitive pressure gauge. The stay of T. Vuleti\'c at LPS, Orsay was in part 
supported by the project {\em R\'eseau formation recherche} of Ministry of 
Education and Research of the Republic of France.
\end{acknowledgement}

\end{document}